\documentclass[conference]{IEEEtran}
\IEEEoverridecommandlockouts

\usepackage{cite}
\usepackage{amsmath,amssymb,amsfonts}
\usepackage{algorithmic}
\usepackage{graphicx}
\usepackage{textcomp}
\usepackage{xcolor}
\def\BibTeX{{\rm B\kern-.05em{\sc i\kern-.025em b}\kern-.08em
    T\kern-.1667em\lower.7ex\hbox{E}\kern-.125emX}}

\begin{document}

\title{Multi-Connectivity for Indoor Terahertz Communication with Self and Dynamic Blockage}

\author{\IEEEauthorblockN{Akram Shafie}
\IEEEauthorblockA{\textit{Research School of Electrical, Energy} \\ \textit{and Materials Engineering} \\
\textit{Australian National University}\\
Canberra, Australia\\
akram.shafie@anu.edu.au}
\and
\IEEEauthorblockN{Nan Yang}
\IEEEauthorblockA{\textit{Research School of Electrical, Energy} \\ \textit{and Materials Engineering} \\
\textit{Australian National University}\\
Canberra, Australia\\
nan.yang@anu.edu.au}
\and
\IEEEauthorblockN{Chong Han}
\IEEEauthorblockA{\textit{University of Michigan--Shanghai}\\
\textit{Jiao Tong University Joint Institute} \\
\textit{Shanghai Jiao Tong University}\\
Shanghai, China\\
chong.han@sjtu.edu.cn}\vspace{-5mm}}
\maketitle

\begin{abstract}
We derive new expressions for the connection probability and the average ergodic capacity to evaluate the performance achieved by multi-connectivity (MC) in an indoor ultra-wideband terahertz (THz) communication system. In this system, the user is affected by both self-blockage and dynamic human blockers. We first build up a three-dimensional propagation channel in this system to characterize the impact of molecular absorption loss and the shrinking usable bandwidth nature of the ultra-wideband THz channel. We then carry out new performance analysis for two MC strategies: 1) Closest line-of-sight (LOS) access point (AP) MC (C-MC), and 2) Reactive MC (R-MC). With numerical results, we validate our analysis and show the considerable improvement achieved by both MC strategies in the connection probability. We further show that the C-MC and R-MC strategies provide significant and marginal capacity gain relative to the single connectivity strategy, respectively, and increasing the number of the user's associated APs imposes completely different affects on the capacity gain achieved by the C-MC and R-MC strategies. Additionally, we clarify that our analysis allows us to determine the optimal density of APs in order to maximize the capacity gain.
\end{abstract}


\section{Introduction}\label{sec:introduction}

Terahertz (THz) communication has been envisioned as a highly promising paradigm to support wireless data applications which demand ultra-high-speed transmission \cite{2018_M6}. These applications, such as wireless virtual reality, augmented reality, and ultra-fast wireless local-area-networks, are beyond the reach of millimeter wave (mmWave) communication. This undoubtedly drives the need of communication at 0.1-10 THz band. Built on the major progress in THz hardware design and THz communication standardization over the past decade, it is anticipated that indoor THz communication systems will be brought to reality in the near future \cite{2019NN6}.

Despite the promise, designing ready-to-use THz communication systems brings new and pressing challenges that have never been seen at lower frequencies \cite{ThzSurvey1}. For example, THz signal propagation suffers from very high spreading loss and molecular absorption loss \cite{THzChan1}. Particularly, the latter is highly frequency-selective and divides the THz band into multiple ultra-wideband transmission windows. Notably, the bandwidth of each transmission window shrinks with longer transmission distance \cite{HBM1}. Moreover, the THz signal propagation is highly vulnerable to blockage, including the blockage caused by the user itself, moving humans, and inherent indoor constructions (e.g., walls and furniture) \cite{2017N3}. All such factors lead to unique propagation characteristics at the THz band, which mandates the design and development of new communication and signal processing mechanisms.

One promising solution to addressing the reliability degradation caused by blockage in THz communication systems is to use multi-connectivity (MC) strategies. Under MC, users are allowed to maintain dynamic association with available access points (APs) for ensuring user session continuity. Due to its importance, the impact of MC on the performance of mmWave communication systems has been examined in recent studies, e.g., \cite{MC4,MC1,MC2,MC3}. Particularly, if the reactive MC (R-MC) strategy is adopted, where the user switches its communication from the current AP to another AP only when the current AP is blocked, has found considerable improvement in the outage probability and capacity \cite{MC2,MC3}. However, given the fundamental difference between mmWave channels and THz channels, the feasibility of using the R-MC strategy in THz communication has not been investigated, which is one of the motivations of this work.

In this paper, we present new analysis to evaluate the impact of MC strategies on the performance of an indoor ultra-wideband THz communication system where the user equipment (UE) suffers from both self-blockage and dynamic human blockage. For this system, we establish a three-dimensional (3D) propagation model where we consider both the spreading loss, determined by 3D propagation distances, and the molecular absorption loss, reflecting the shrinking usable bandwidth nature of the ultra-wideband THz channel. Under such consideration, we derive new expressions for the connection probability and the average ergodic capacity for two MC strategies, namely, closest line-of-sight (LOS) AP MC (C-MC) strategy and R-MC strategy. Different from the R-MC strategy, the UE under the C-MC strategy always communicates with the closest LOS AP, while maintaining association with several APs. Here, the connection probability is defined as the probability that at least one associated AP is LOS such that the UE can connect and communicate with. Aided by numerical results, we demonstrate that our analysis is accurate and a considerably improved connection probability is achieved by MC strategies, relative to the single connectivity (SC) strategy. Moreover, we find that the C-MC strategy achieves a significantly higher capacity gain over the SC strategy than the R-MC strategy. Furthermore, we reveal that when the UE is associated with more APs, the capacity gain achieved by the C-MC strategy increases while that achieved by the R-MC strategy decreases, sometimes even below zero. This worsening capacity behavior for the R-MC strategy demonstrates the impracticality of using it for THz communication, especially at low density of APs, which is different from the conclusion for mmWave communication. Finally, we clarify that there is an optimal density of APs to maximize the capacity gain of MC strategies, the value of which can be determined by using our analysis.

\section{System Model}\label{sec:system_model}

\begin{figure}[t]
    \centering
    \includegraphics[width=0.82\columnwidth]{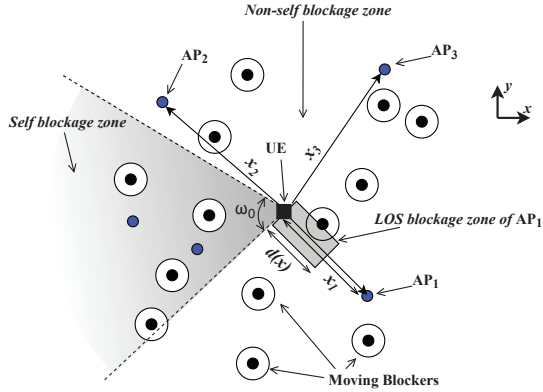}
    \caption{Top view of a 3D THz communication system where a UE associates with non-blocked APs.}\label{Fig:SysModel1}\vspace{-2mm}
\end{figure}

\begin{figure}[t]
    \centering
    \includegraphics[width=0.6\columnwidth]{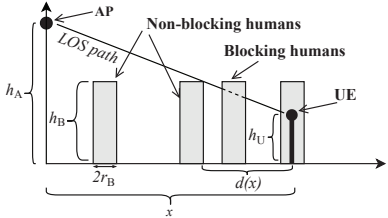}
    \caption{Side view of a single AP-UE link in the considered 3D THz communication system.}\label{Fig:Ng2}\vspace{-4mm}
\end{figure}

In this work, we consider a 3D THz communication system, as depicted in Figs.~\ref{Fig:SysModel1} and \ref{Fig:Ng2}. Specifically, Fig.~\ref{Fig:SysModel1} shows the top view of the considered system where a UE associates with THz APs that are not blocked by the UE itself nor by moving humans, while Fig.~\ref{Fig:Ng2} shows the side view of a single link from an AP to a UE. The UE is of height $h_{\textrm{U}}$ and assumed to be stationary. The APs are of fixed height $h_{\textrm{A}}$ and their location follows a Poisson point process (PPP) in $\mathbb{R}^{2}$ with the density of $\lambda_{\textrm{A}}$. Moving humans in the area of interest acts as potential blockers. These humans are modeled as cylinders with the radius of $r_{\textrm{B}}$ and height of $h_{\textrm{B}}$ and their location follows another PPP with the density of $\lambda_{\textrm{B}}$. Considering the practicality of THz communication system, we assume that $h_{\textrm{A}} > h_{\textrm{B}} > h_{\textrm{U}}$.

We assume that the mobility of humans follows the random directional model (RDM). Based on this model, a moving human randomly selects a direction to travel in and a time duration for this travel \cite{RDM1}. Similar to \cite{MC3}, in this work we assume that the moving speed is $v_{\textrm{B}}$. If a blocker is moving as per the RDM model in a given area in  $\mathbb{R}^2$, the probability density function (PDF) of the location of blockers is uniform over time \cite{RDM1}. As such, at any given time instant, the location of blockers forms PPP with the same density, $\lambda_{\textrm{B}}$.

\subsection{Propagation Model}

\begin{figure}[t]
    \centering
    \includegraphics[width=0.9\columnwidth]{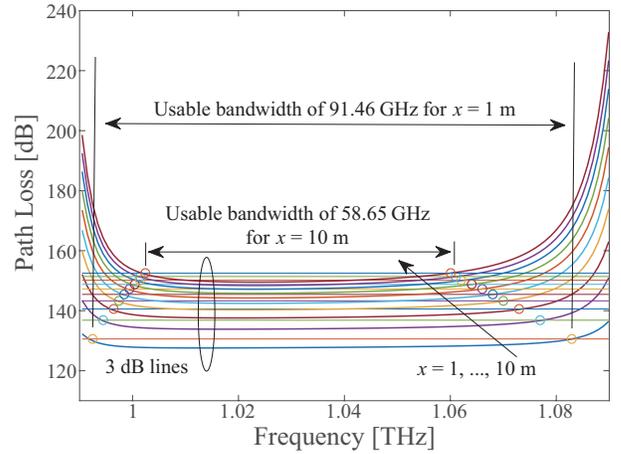}
    \caption{Usable bandwidth as a function of distance \cite{HBM1}.}
    \label{Fig:3dBband}\vspace{-4mm}
\end{figure}

The signal propagation at THz frequencies is determined by spreading loss and molecular absorption loss \cite{THzChan1}. The path loss of an arrival ray in the 3D THz channel is given by
\begin{equation}\label{Equ:PLTHz}
L(f,x)=\left(\frac{4 \pi f \overline{x}}{c}\right)^{2}e^{K_{\textrm{abs}}(f)\overline{x}},
\end{equation}
where $f$ is the operating frequency, $c$ is the speed of light, $x$ and $\overline{x}$ are the 2D and 3D propagation distances between the UE and the AP, respectively, with $\overline{x} = \sqrt{(h_{\textrm{A}} - h_{\textrm{U}})^2 +x^2}$, and $K_{\textrm{abs}}(f)$ is the frequency-dependent absorption coefficient of the transmission medium. Here, $\left({4 \pi f \overline{x}}/{c}\right)^{2}$ represents spreading loss and $e^{K_{\textrm{abs}}(f)\overline{x}}$ represents molecular absorption loss.
In this work, we use the absorption coefficient values which are calculated according to \cite{2011_Jornet_TWC} for the standard atmosphere with $10\%$ humidity.

The impact of molecular absorption loss on the THz band is two-fold. First, as shown in \cite[Fig. 1]{HBM1}, the whole THz band is divided into multiple ultra-wideband transmission windows, due to the intermittent absorption loss peaks which are observed throughout the THz band at different frequencies. Second, path loss varies drastically even within a specific ultra-wideband transmission window, and this variation further increases when transmission distance increases. As a result, the usable bandwidth within the transmission window of interest becomes narrower for longer transmission distance. In this work, we refer to the \emph{usable bandwidth} as the bandwidth where the path loss variation is within 3 dB. Fig. \ref{Fig:3dBband} shows the path loss corresponding to one transmission window ($0.99-1.09~\textrm{THz}$) for different transmission distances. Specifically, the available bandwidth drops from $91.46~\textrm{GHz}$ to $58.65~\textrm{GHz}$ when the transmission distance increases from $1~\textrm{m}$ to $10~\textrm{m}$.

In this work, we concentrate on one ultra-wideband transmission window when THz communication occurs. Considering the aforementioned distance varying nature of the usable bandwidth within an ultra-wideband transmission window, we assume that both the UE and the AP choose the appropriate usable bandwidth according to the transmission distance. Also, through utilizing the usable bandwidth, instead of the total transmission window, at different transmission distances, the broadening effect of broadband signals can be restricted within reasonable limit \cite{A0}. Moreover, in this work we focus on the LOS rays of THz signals. This is because when signals are propagated at THz band frequencies, the direct ray dominates the received signal energy, due to the high directional nature and the high reflection loss of THz beams \cite{A0}. Furthermore, in this work we assume that the considered system is noise-limited such that the interference from other APs is not considered. We clarify that this is a valid assumption in 3D THz communication systems since both the interfering probability and the power from interfering APs are minimal, due to the high directionality of APs and the extremely high path loss in THz transmission.

\subsection{Blockage}

THz waves are highly susceptible to blockages and can be blocked even by the UEs themselves. In our system, we consider that the blockage of an AP-UE link is caused by either the UE itself, referred to as \emph{self-blockage}, or the dynamic human blockers.

\subsubsection{Self-Blockage}

Self-blockage plays a significant role in determining THz system performance. Notably, self-blockage may lead to the fact that some APs surrounding a UE are totally inaccessible, even if they are within close proximity. Against this background, we define the zone which is not blocked by the UEs themselves as ``non-self-blockage zone'' \cite{mmM2}, as shown in Fig. \ref{Fig:SysModel1} with a non-self-blockage angle of $\omega=2\pi-\omega_{0}$. We consider that the UE only associates with the APs which are located in its non-self-blockage zone.

\subsubsection{Dynamic Human Blockage}

The LOS link between an AP and the UE is blocked if at least one blocker appears in the LOS blockage zone of the AP-UE link. This area can be approximated by a rectangle between the UE and the AP with sides of $2r_{\textrm{B}}$ and $d(x)$, as shown in Figs. \ref{Fig:SysModel1} and \ref{Fig:Ng2}, where
\begin{equation}\label{Equ:dx}
d(x)=\frac{h_{\textrm{B}}-h_{\textrm{U}}}{h_{\textrm{A}}-h_{\textrm{U}}}x+r_{\textrm{B}}.
\end{equation}
Therefore, the LOS probability of an AP-UE link with a 2D distance $x$ is the same as the void probability of Poisson process in the LOS blockage zone, which is given by
\begin{equation}\label{Equ:PL1}
p_{\textrm{L}}(x)= e^{-2 r_{\textrm{B}} \lambda_{\textrm{B}} d(x)}	
= \zeta  e^{-\beta x},
\end{equation}
where $\zeta  = e^{-2\lambda_{\textrm{B}}r_{\textrm{B}}^2}$ and $\beta = 2\lambda_{\textrm{B}}r_{\textrm{B}}(h_{\textrm{B}}-h_{\textrm{U}})/(h_{\textrm{A}}-h_{\textrm{U}})$.

Apart from the LOS probability, we are also interested in the statistics of time duration within which a specific AP remains blocked or non-blocked, which is necessary for our analysis under different MC strategies. To this end, as explained in \cite{MC3}, the time duration that an AP remains blocked or non-blocked is modeled by an alternating renewal process, where $t_{\textrm{LOS}}$ and $t_{\textrm{NLOS}}$ denote the random variables characterizing a non blockage and blockage time duration, respectively \cite{RenProc}. Since blockers enter the LOS blockage zone according to a Poisson process, the time duration an AP remains unblocked follows an exponential distribution with the temporal density of $\mu_{\textrm{B}}(x)$, where $\mu_{\textrm{B}}(x) = 2 r_{\textrm{B}} v_{\textrm{B}}\lambda_{\textrm{B}}d(x)$ \cite{MC3}. Accordingly, the mean of a single non blockage time duration is calculated as $\mathbb{E}[t_{\textrm{LOS}};x]=1/\mu_{\textrm{B}}(x)$.

It is noted that the LOS probability $p_{\textrm{L}}(x)$ in \eqref{Equ:PL1} can be interpreted as the fraction of time that the UE is connected with the AP at the distance $x$. Therefore, for a given AP-UE distance $x$, $p_{\textrm{L}}(x)$ can be re-expressed as
\begin{equation}\label{Equ:PL2}
p_{\textrm{L}}(x)=\frac{\mathbb{E}[t_{\textrm{LOS}};x]}
{\mathbb{E}[t_{\textrm{LOS}};x]+\mathbb{E}[t_{\textrm{NLOS}};x]}.
\end{equation}
Thereafter, by jointly considering \eqref{Equ:PL1} and \eqref{Equ:PL2}, the mean of a single blockage time duration, $\mathbb{E}[t_{\textrm{NLOS}};x]$, is expressed as
\begin{equation}\label{Equ:ENLOS}
\mathbb{E}[t_{\textrm{NLOS}};x]=\frac{1-\zeta  e^{-\beta x}}{2 \zeta r_{\textrm{B}}v_{\textrm{B}}\lambda_{\textrm{B}}d(x)e^{-\beta x}}.
\end{equation}

\subsection{Connectivity Strategies}

In this work, we investigate two $N$-degree MC strategies while considering the SC strategy as the benchmark. Under the SC strategy, the UE only associates with its closest AP in its non-self-blockage zone. As such, if the LOS link between the UE and its closest AP is blocked, the UE is in outage. Different from the SC strategy, the UE under $N$-degree MC strategies selects the $N$ closest APs in its non-self-blockage zone and associates with them for user session continuity. At any given time instant, even if more than one out of such $N$ APs are in LOS, the UE only connects and communicates with one LOS AP. At the same time, the UE maintains the rest of the LOS APs as active backup APs to enable instantaneous AP re-association with negligible switching time, whenever the currently connected AP is blocked.

Our considered two MC switching strategies are as follows:
\begin{itemize}
\item
C-MC: Under this strategy, out of the $N$ associated APs, the UE always connects and communicates with the closest LOS AP at any time instant.
\item
R-MC: Under this strategy, the UE switches its connection and communication from the current AP to another AP, only when the current AP is blocked. By the end of this switching, the UE connects and communicates with the closest LOS AP out of the $N$ associated APs.
\end{itemize}
We note that the C-MC strategy can provide the best performance but lead to frequent AP switching, while the R-MC strategy is a ``lightweight'' solution in terms of software and hardware implementations.



\subsection{Distance Distribution of APs}

In this subsection, by considering self-blockage, we derive the conditional joint PDF of the distances to $N$ closest APs. This result will be used in the next section to determine the connection probability and the ergodic capacity of MC strategies. In this derivation, we consider the APs are located farther than $R_{0}$ from the UE, e.g., $R_{0}=1~\textrm{m}$ for indoor THz systems. This consideration is necessary to bring the benefits of MC strategies to the considered system. Indeed, if the distances between the UE and some APs are short, e.g., less than $R_{0}$, the LOS blockage zone is very small, which may lead to the fact that such APs are LOS always and there is no need for the UE to switch its communication among APs.

Let AP$_i$ denote the $i$th closest AP from the UE and $x_{i}$ denote the distance between AP$_i$ and UE, where $i\in\{1,\cdots,N\}$. The conditional PDF of the distance from the UE to the closest AP, i.e., AP$_{1}$, is given by
\begin{equation}\label{Equ:PDF1}
f(x_{1})=\omega\lambda_{\textrm{A}}x_{1}e^{\frac{\omega}{2}\lambda_{\textrm{A}}R_{0}^2}
e^{-\frac{\omega}{2}\lambda_{\textrm{A}}x_{1}^2}.
\end{equation}
The conditional PDF of the distance from the UE to AP$_2$, given that AP$_1$ is at distance $x_{1}$, is given by
\begin{equation}\label{Equ:PDF2}
f(x_{2}|x_{1})=\omega\lambda_{\textrm{A}}x_{2}e^{-\frac{\omega}{2}\lambda_{\textrm{A}}\left(x_{2}^2-x_{1}^2\right)}.
\end{equation}
Using \eqref{Equ:PDF1} and \eqref{Equ:PDF2}, the conditional joint PDF of the distances from the UE to the closest and second closest APs, i.e., AP$_{1}$ and AP$_{2}$, is derived as
\begin{align}\label{Equ:PDF3}
f (x_{1},x_{2})&=f(x_{2}|x_{1})f(x_{1})\notag\\
&=(\omega\lambda_{\textrm{A}})^2 x_{1} x_{2} e^{\frac{\omega}{2}\lambda_{\textrm{A}}R_{0}^2}
e^{-\frac{\omega}{2} x_{2}^2}.
\end{align}
Thereafter, by reapplying the procedure used for deriving $f(x_{1},x_{2})$ from $f(x_{1})$ for $N-1$ times and continuing along the line, the joint PDF of AP$_{1}$ to AP$_{N}$, denoted by $f(x_{1},x_{2},\cdots, x_{N})$, is derived as
\begin{equation}\label{Equ:PDF4}
f (x_{1},x_{2},\cdots, x_{N})=(\omega\lambda_{\textrm{A}})^{N}
e^{\frac{\omega}{2}\lambda_{\textrm{A}}R_{0}^2}
\prod_{i=1}^{N}x_{i}e^{-\frac{\omega}{2}x_{N}^2}.
\end{equation}

\section{Analysis of Connection Probability and Ergodic Capacity}\label{sec:analysis}

\subsection{Connection Probability}

\subsubsection{Single Connectivity}

For the SC strategy, as the UE only associates with its closest AP, connection cannot be established if the LOS link between the UE and its closest AP is blocked. As such, for an AP-UE distance of $x_{1}$, the connection probability at the UE is given by
\begin{equation}\label{Equ:pcSC1}
p_{\textrm{c}}(x_{1})=p_{\textrm{L}}(x_{1})=\zeta  e^{-\beta x_{1}}.
\end{equation}
Considering that the location of APs follows a PPP, the average connection probability for the SC strategy is derived as
\begin{align}\label{Equ:pcSC2}
p_{\textrm{c},\textrm{SC}}&=\mathbb{E}_{x_{1}}\left[p_{\textrm{c}}(x_{1})\right]
=\int_{R_{0}}^{\infty}p_{\textrm{c}}(x_{1}) f (x_{1}) dx_{1}\notag\\
&=\zeta  e^{-\beta R_{0}}\left[1-\beta\sqrt{\frac{\pi}{2\lambda_{\textrm{A}}\omega}}e^{\varpi^{2}}
\textrm{erfc}\left(\varpi\right)\right],
\end{align}
where $\textrm{erfc}(\cdot)$ denotes the cumulative error function and $\varpi=(\beta+\lambda_{\textrm{A}}R_{0}\omega)/\sqrt{2\lambda_{\textrm{A}}\omega}$ \cite{IntegralBook}.

\subsubsection{Multi-Connectivity}

Considering that the blockage process of each link is independent and AP$_{1}$ $\cdots$ AP$_{N}$ are at distances $x_{1},\cdots, x_{N}$ from the UE, respectively, the connection probability at the UE is given by
\begin{align}\label{Equ:pcMC1}
&p_{\textrm{c}}(x_{1},x_{2},\cdots,x_{N})=p_{\textrm{L}}(x_{1},x_{2},\cdots,x_{N})\notag\\
&=1-\prod_{i=1}^{N}\left(1 - p_{\textrm{L}}(x_{i})\right)
=1-\prod_{i=1}^{N}(1 - \zeta  e^{-\beta x_{i}}).
\end{align}
Therefore, the average connection probability for the $N$-degree MC strategy is derived as
\begin{align}\label{Equ:pcMC2}
p_{\textrm{c},\textrm{MC}}
&=\int_{R}^{\infty}\int_{x_{1}}^{\infty}\cdots\int_{x_{N-1}}^{\infty}
\left(1-\prod_{i=1}^{N}\left(1 - \zeta  e^{-\beta x_{i}}\right)\right)\notag\\
&\hspace{2mm}\times\left(\omega\lambda_{\textrm{A}}\right)^{N}e^{\frac{\omega}{2}\lambda_{\textrm{A}}R_{0}^2} \prod_{i=1}^{N}x_{i}e^{-\frac{\omega}{2}x_{N}^2} dx_{N} \cdots dx_{1},
\end{align}
which can be calculated numerically. It is noted that in our considered system, the two $N$-degree MC strategies achieve the same connection probability since switching time is assumed to be negligible during AP re-association.

\subsection{Ergodic Capacity}

\subsubsection{Single Connectivity}

Due to the frequency dependant nature of the THz wideband, we decompose the received signal power as the sum of powers of sub-bands, where each sub-band channel is narrow and has a flat band response. Hence, the wideband capacity when the UE is connected and communicated with an AP at distance $x$ is calculated as the sum of capacity of each sub-band, which results in
\begin{equation}\label{Equ:CapTHz}
C(x)=\!\! \sum_{\eta=1}^{N_{\textrm{B}}(x)}\!\!\Delta f \log\left(1{+}\frac{P_{\textrm{T},\eta}(x) G_{\textrm{A}} G_{\textrm{U}} L^{{-}1}(f_{\eta},x) }{\Delta f N_{0}}\right),
\end{equation}
where $N_{\textrm{B}}(x)$ is the total number of sub-bands within the usable bandwidth corresponding to distance $x$, $\Delta f = 1~\textrm{GHz}$ is the width of each sub-band, $P_{\textrm{T},\eta}(x)$ is the transmit power in the $\eta$th sub-band with the total transmit power $P_{\textrm{T}} =\sum_{\eta=1}^{N_{\textrm{B}}(x)} P_{\textrm{T},\eta}(x)$ being fixed, $G_{\textrm{A}}$ and $G_{\textrm{U}}$ are the antenna gains at the AP and the UE, respectively, $N_{0}$ is the additive white Gaussian noise power, and $L^{-1}(f_{\eta},x)$ is given by \eqref{Equ:PLTHz}.

It is noted that as the AP and the UE utilize distance-aware bandwidth adaptation, $N_{\textrm{B}}(x)$ is different from one transmission distance to another. Thus, using \eqref{Equ:CapTHz} and the blockage probability in \eqref{Equ:pcSC1}, we derive the average ergodic capacity for the SC strategy as
\begin{align}\label{Equ:CSC1}
C_{\textrm{SC}}&{=}\mathbb{E}_{x_{1}}\left[p_{\textrm{c}}(x_{1})C(x_{1})\right]
=\!\!\int_{R_{0}}^{\infty}\!\!\!p_{\textrm{c}}(x_{1})C(x_{1})f(x_{1})dx_{1},
\end{align}
which can be calculated numerically.

\subsubsection{Multi-Connectivity}

Given that AP$_{1}$ $\cdots$ AP$_{N}$ are at distances $x_{1},\cdots, x_{N}$ from the UE, respectively, the ergodic capacity of the $N$-degree $\Psi$-MC strategy, where $\Psi\in\{\textrm{C},\textrm{R}\}$, is written as
\begin{equation}\label{Equ:CapMC1}
C_{\Psi{-}\textrm{MC}}^{N}\left(x_{1},\cdots\!,x_{N}\right)\!=\!
p_{\textrm{L}}\left(x_{1},\cdots\!,x_{N}\right)\!\! \left[\prod_{i=1}^{N}\!\gamma_{\Psi,i}C(x_{i})\right]\!,
\end{equation}
where $\gamma_{\Psi,i}$ represents the percentage of time that the UE is connected to AP$_{i}$ within the total non-outage duration when the $\Psi$-MC strategy is utilized. We clarify that the ergodic capacity for the C-MC strategy is different from that for the R-MC strategy. This is due to the fact that within the total non-outage time duration, the percentage of time that the UE is connected with a particular AP for one strategy is different from the other, i.e., $\gamma_{\textrm{C},i}\neq\gamma_{\textrm{R},i}$.

Using \eqref{Equ:CapMC1}, the average ergodic capacity for the $\Psi$-MC strategy is written as
\begin{align}\label{Equ:CapMC2}
C_{\Psi-\textrm{MC}}^{N}&=\int_{R}^{\infty}\int_{x_{1}}^{\infty}\cdots\int_{x_{N-1}}^{\infty} C_{\Psi-\textrm{MC}}^{N}(x_{1},\cdots,x_{N})\notag\\
&\hspace{5mm}\times f\left(x_{1},\cdots,x_{N}\right) dx_{N} \cdots dx_{1}.
\end{align}
We next derive $\gamma_{\Psi,i}$ for the two $N$-degree MC strategies.

\begin{figure}[t]
    \centering
    \includegraphics[scale=0.16]{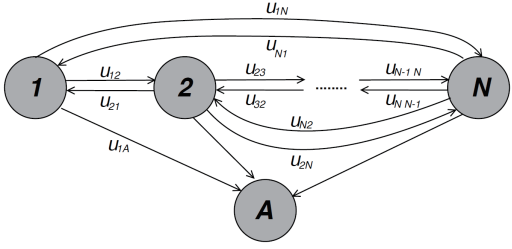}
    \caption{The absorbing Markov chain model for the reactive AP switching process.}\label{fig:my_label2}
\end{figure}

$\triangleright$~\textit{C-MC:} The RDM model indicates that the blockers moving according to this model in a certain area is distributed uniformly in this area \cite{RDM1}. As such, the percentage of the time a UE stays connected with AP$_{i}$ within the total non-outage duration for the C-MC strategy, $\gamma_{\textrm{C},i}$, is given by
\begin{equation}\label{Equ:gamCLOS}
\gamma_{\textrm{C},i}=\frac{p_{\textrm{L}}(x_{i})\prod_{j=1}^{i-1}\left(1-p_{\textrm{L}}(x_{j})\right)}
{p_{\textrm{L}}\left(x_{1},x_{2},\cdots,x_{N}\right)}.
\end{equation}
By substituting \eqref{Equ:gamCLOS} into \eqref{Equ:CapMC1} and using \eqref{Equ:CapMC2}, the average ergodic capacity of C-MC strategy is obtained.

$\triangleright$~\textit{R-MC:} In this strategy, the UE first communicates with its closest LOS AP, out of the $N$ associated APs, and switches to another when the current AP is blocked. This switching process continues in a reactive manner among the $N$ APs until the time instant where all the $N$ APs are blocked, which leads to outage. While waiting in outage, once an AP comes in LOS, the UE starts to communicate with this AP; from there onwards, the switching pattern continues as aforementioned. As indicated in \cite{MC3}, this AP switching process can be modeled as an absorbing Markov chain which is depicted in Fig. \ref{fig:my_label2}, where state $i$ represents $i$th closest AP and the absorbing state, \textit{state A}, represents the outage when all the $N$ APs are blocked.

To parameterize the absorbing Markov chain in Fig.~\ref{fig:my_label2}, it is necessary to identify the matrix $\mathbf{U}$ that contains the transition probabilities between the transient states ${1,\cdots,N}$ and the initial state vector $\mathbf{b}$. Given that AP$_{1}$ $\cdots$ AP$_{N}$ are at distances $x_{1},\cdots, x_{N}$ from the UE, respectively, the element in the $i$th row and $j$th column of $\mathbf{U}$, where $i\in\{1,\cdots,N\}$ and $j\in\{1,\cdots,N\}$, is given by
\begin{equation}\label{Equ:AMC1}
u_{ij}=
\begin{cases}
p_{\textrm{L}}\left(x_{j}\right)\prod_{k=1}^{j-1}\left(1-p_{\textrm{L}}\left(x_{k}\right)\right), & \mbox{if $i\neq j$}, \\
0, & \mbox{otherwise}.
\end{cases}
\end{equation}

As explained in \cite{MC3}, it is cumbersome to derive an exact expression for the elements of $\mathbf{b}$, since the AP that initiates the chain following the absorbing state depends on the last AP of the previous chain. By disregarding this dependency and establishing that $b_{i}$ is proportional to the mean duration of the blockage period \cite{MC3}, the $i$th element in $\mathbf{b}$ is given by
\begin{equation}\label{Equ:AMC2}
b_{i}=\frac{\mathbb{E}\left[t_{\textrm{NLOS}};x_{i}\right]}
{\sum_{j=1}^{N}\mathbb{E}\left[t_{\textrm{NLOS}};x_{j}\right]},~i\in\{1,\cdots,N\}.
\end{equation}

\begin{table}[!t]
\caption{Value of System Parameters Used in Section~\ref{sec:numerical}}\vspace{-2mm}
\begin{center}
\begin{tabular}{|c|c|c|}
\hline
\textbf{Parameter} & \textbf{Symbol}& \textbf{Value} \\
\hline
Height of APs and UE & $h_{\textrm{A}}$, $h_{\textrm{U}}$  & $3.0 $ m, $1.2 $ m\\
\hline
Height and radius of blockers & $h_{\textrm{B}}, r_{\textrm{B}}$ & $1.7 $ m, $0.3 $ m\\
\hline
Antenna gains & $G_{\textrm{A}}$, $G_{\textrm{U}}$  & $25~\textrm{dBi}$, $25~\textrm{dBi}$  \\
\hline
Non self-blockage angle & $\omega$  & $\pi$ \\\hline
Speed of blockers & $v_{\textrm{B}}$  & 1 ms$^{-1}$ \\ \hline
Blocker density & $\lambda_{\textrm{B}}$  &  0.2  m$^{-2}$ \\ \hline
Transmission windows & $W_{1}$ & $0.99-1.09~\textrm{THz}$, \\
& $W_{2}$ & $3.34-3.49~\textrm{THz}$\\\hline
Transmit Power & $P_{\textrm{T}}$ &  $20~\textrm{dBm}$,  $30~\textrm{dBm}$ \\\hline
\end{tabular}\label{tab1}
\end{center}\vspace{-2mm}
\end{table}

We note that $\gamma_{\textrm{R},i}$ can be characterized by the mean of times that the absorbing Markov chain visits a transient state $i$ before reaching the absorbing state. Based on the Markov chain theory~\cite{AMC}, this mean can be determined by the elements of the fundamental matrix $\mathbf{D}=\left(\mathbf{I}-\mathbf{U}\right)^{-1}$ and the initial state vector $\mathbf{b}$. Therefore, using \eqref{Equ:AMC1} and \eqref{Equ:AMC2} and rectifying the oversight in \cite[Eq (62)]{MC3}, we derive $\gamma_{R,i}$ as
\begin{equation}\label{Equ:gamReact}
\gamma_{\textrm{R},i}=\sum_{j=1}^{N}\frac{b_{j}d_{j,i}\mathbb{E}\left[t_{\textrm{LOS}};x_{i}\right]}
{\sum_{k=1}^{N}d_{j,k}\mathbb{E}[t_{\textrm{LOS}};x_{k}]}.
\end{equation}
Finally, by substituting \eqref{Equ:gamReact} into \eqref{Equ:CapMC1} and using \eqref{Equ:CapMC2}, the average ergodic capacity of R-MC strategy is obtained.

\section{Numerical Results and Discussion}\label{sec:numerical}

In this section, we present numerical results to evaluate the impact of MC strategies and parameters on the performance of the considered system with SC being the benchmark. The values of the parameters used in this section are summarized in Table \ref{tab1}, unless specified otherwise.

\begin{figure}[!t]
    \centering
    \includegraphics[height=2.19in,width=0.94\columnwidth]{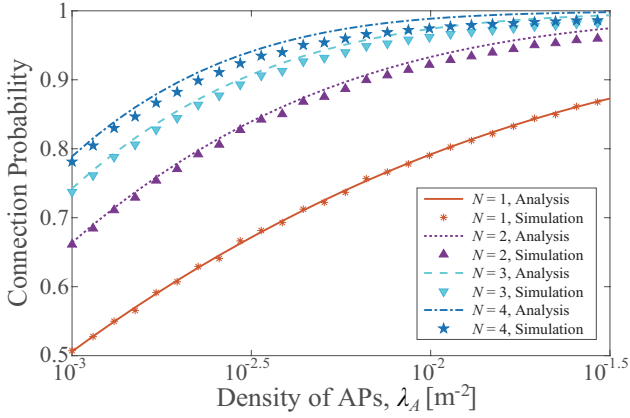}
    \caption{Connection probabilities of SC and MC strategies versus the density of APs.}\label{Fig:NR1}
\end{figure}

\begin{figure}[t]
    \centering
    \includegraphics[height=2.19in,width=0.94\columnwidth]{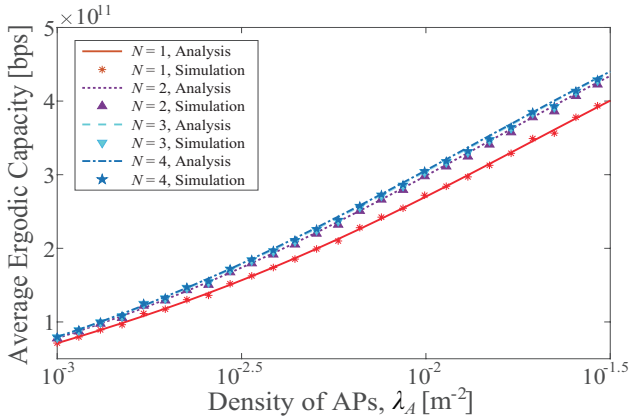}
    \caption{Average ergodic capacity of C-MC strategy versus the density of APs for the transmission window of $0.99-1.09~\textrm{THz}$ with $P_{\textrm{T}}=20~\textrm{dBm}$.}\label{Fig:NR2}
\end{figure}

Fig.~\ref{Fig:NR1} plots the connection probabilities of the SC strategy with $N=1$ and the $N$-degree MC strategy with $N=2$, $3$, and $4$ versus the density of APs, $\lambda_{\textrm{A}}$. We first observe that the connection probability significantly increases when $N$ becomes higher. This demonstrates the reliability performance improvement brought by the MC strategy relative to the SC strategy. Second, we observe the profound increase in the connection probability when $\lambda_{\textrm{A}}$ is larger. This implies that a denser deployment of APs can effectively overcome the detrimental impact of self-blockage and dynamic human blockage. Notably, when $\lambda_{\textrm{A}}$ is large, e.g., $\lambda_{\textrm{A}}=1.5\times10^{-2}$, it is possible to achieve a high connection probability, e.g., $95\%$, even for $N=2$. Third, we observe that our analysis well match the simulations, especially for the SC strategy with any $\lambda_{\textrm{A}}$ and for the MC strategy with low and medium $\lambda_{\textrm{A}}$, which demonstrates the correctness of our analysis. When $\lambda_{\textrm{A}}$ is high, our analysis for the MC strategy slightly overestimate the connection probability. This is due to the fact that our analysis is under the assumption that the blockage process from different APs are independent of each other. However, for high $\lambda_{\textrm{A}}$, non-negligible dependencies appear such that the LOS blockage zones of different AP-UE links overlap with each other, which yields the slight overestimation.

Fig. \ref{Fig:NR2} plots the average ergodic capacity of the C-MC strategy versus $\lambda_{\textrm{A}}$ considering the transmission window of $0.99-1.09~\textrm{THz}$ and the total transmit power of $P_{\textrm{T}}=20~\textrm{dBm}$. As expected, the average ergodic capacity becomes noticeably higher when $N$ increases from $1$ to $2$, which shows the benefits of the MC strategy relative to the SC strategy. Moreover, we note that the average ergodic capacity gain brought by further increasing $N$ from $2$ to $4$ is marginal. This marginal gain in capacity, along with the profound increase in connection probability when $\lambda_{\textrm{A}}$ is larger, which is observed in Fig. \ref{Fig:NR1}, implies that for the MC strategy, it may not be worthwhile to allow the UE to associate with more than two closest APs at higher AP densities.
Furthermore, we again observe the benefit of a denser deployment of APs and the correctness of our analysis.

\begin{figure}[t]
    \centering
    \includegraphics[height=2.19in,width=0.94\columnwidth]{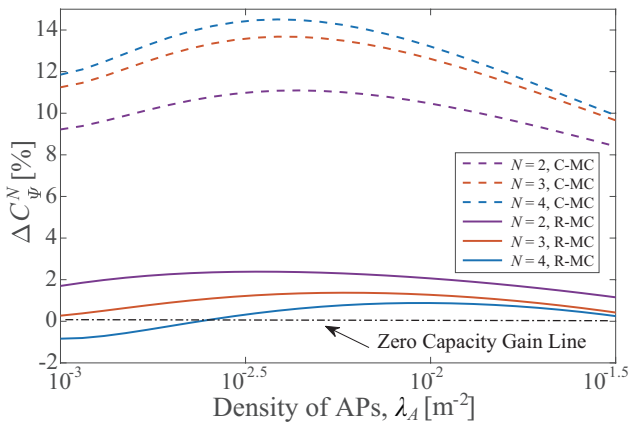}
    \caption{Capacity gain of MC strategy relative to SC strategy versus the density of APs for the transmission window of $0.99-1.09~\textrm{THz}$ with $P_{\textrm{T}}=20~\textrm{dBm}$.}\label{Fig:NR3}
\end{figure}

In order to precisely examine the performance improvement brought by the MC strategies relative to the SC strategy, we now define the capacity gain of the MC strategies relative to the SC strategy as
\begin{equation}
{\Delta}C_{\Psi}^{N}=\frac{C_{\Psi-\textrm{MC}}^{N}-C_{\textrm{SC}}}{C_{\textrm{SC}}},
\end{equation}
where $\Psi\in\{\textrm{C},\textrm{R}\}$. Fig. \ref{Fig:NR3} plots ${\Delta}C_{\textrm{C}}^{N}$ and ${\Delta}C_{\textrm{R}}^{N}$ versus $\lambda_{\textrm{A}}$ for the transmission window of $0.99-1.09~\textrm{THz}$ with $P_{\textrm{T}}=20~\textrm{dBm}$. Moreover, the zero capacity gain is plotted in this figure. We first observe that relative to the SC strategy, the C-MC strategy achieves a large capacity gain while the R-MC strategy achieves a small capacity gain, e.g., ${\Delta}C_{\textrm{C}}^{N}\approx10\%$ while ${\Delta}C_{\textrm{R}}^{N}\approx2\%$ when $\lambda_{\textrm{A}} = 1.5\times10^{-2}$ and $N=2$. This observation indicates that the C-MC strategy significantly outperforms the R-MC strategy in THz communication systems. This observation is due to the fact that under the C-MC strategy, the UE always communicates with the closed LOS AP but under the R-MC strategy, the UE may communicate with a farther LOS AP. Indeed, the distance is a key factor governing the performance of THz communication systems given that it narrows the usable bandwidth and increases molecular absorption loss. Second, we observe that when $N$ increases, ${\Delta}C_{\textrm{C}}^{N}$ increases but ${\Delta}C_{\textrm{R}}^{N}$ decreases. The observation for the C-MC strategy is expected since under this strategy, associating with more APs gives the UE more opportunities to maintain the high-capacity communication when the current AP is blocked. Differently, under the R-MC strategy, associating with more APs gives the UE a higher chance to communicate with a farther LOS AP. This leads to the fact that the percentage of time that the UE communicates with the closest LOS AP reduces, thus decreasing the capacity. Third, we observe that for some cases, e.g., $N=4$ and small $\lambda_{\textrm{A}}$, ${\Delta}C_{\textrm{R}}^{N}$ gain is less than 0. This shows the impracticality of using the R-MC strategy for the THz communication system with a low density of APs. Finally, we observe that there exists the optimal density of APs which maximizes the capacity gain for both MC strategies. Notably, this optimal density can be determined by using our analysis.

\begin{figure}[t]
    \centering
    \includegraphics[height=2.19in,width=0.94\columnwidth]{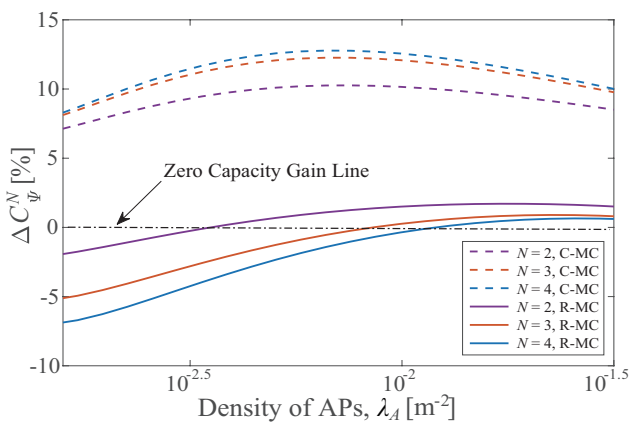}
    \caption{Capacity gain of MC strategy relative to SC strategy,  ${\Delta}C_{\Psi}^{N}$, versus the density of APs for the transmission window of $3.34-3.49~\textrm{THz}$ with $P_{\textrm{T}}=30~\textrm{dBm}$.}\label{Fig:NR4}
\end{figure}

To evaluate the capacity gain within a different transmission window, Fig. \ref{Fig:NR4} plots ${\Delta}C_{\textrm{C}}^{N}$ and ${\Delta}C_{\textrm{R}}^{N}$ versus $\lambda_{\textrm{A}}$ for the transmission window of $3.34-3.49~\textrm{THz}$ with $P_{\textrm{T}}=30~\textrm{dBm}$. Once again, we observe in Fig. \ref{Fig:NR4} that ${\Delta}C_{\textrm{C}}^{N}>{\Delta}C_{\textrm{R}}^{N}$, increasing $N$ increases ${\Delta}C_{\textrm{C}}^{N}$ but decreases ${\Delta}C_{\textrm{R}}^{N}$, and the optimal density of APs maximizes the capacity gain, which is similar to Fig. \ref{Fig:NR3}. Apart from these similar observations, we further observe that although $P_{\textrm{T}}$ increases from $20~\textrm{dBm}$ in Fig. \ref{Fig:NR3} to $30~\textrm{dBm}$ in Fig. \ref{Fig:NR4}, the capacity gains achieved by the C-MC and R-MC strategies are worse when the transmission window increases from $0.99-1.09~\textrm{THz}$ to $3.34-3.49~\textrm{THz}$.
This implies that, using wider transmission windows of higher THz band as compared to transmission windows of lower THz band,  do not yield any additional improvement in capacity gain for MC strategies, due to the increased molecular absorption loss at higher THz frequencies.  In addition, we observe that the capacity gains achieved by the R-MC strategy is negative even for medium $\lambda_{\textrm{A}}$, which is dfferent from Fig. \ref{Fig:NR3}. This is due to the combined effect of the molecular absorption loss which becomes more severe when the transmission frequency increases, and the reactive switching nature of the R-MC strategy. This again shows the impracticability of using the R-MC strategy for higher THz frequencies.

\section{Conclusion}\label{sec:conclusion}

We analyzed the performance achieved by MC strategies which are used to combat the non-connection effect caused by self-blockage and dynamic blockage in THz communication systems. Specifically, we developed a new analytical framework to evaluate the connection probability and the average ergodic capacity achieved by C-MC and R-MC strategies. Using numerical results, we demonstrated the accuracy of our analysis and revealed several insights. First, comparing to the SC strategy, using MC strategy leads to a considerable improvement in the connection probability. Second, the capacity gain brought by the C-MC strategy over the SC strategy is significant, while that brought by the R-MC strategy is marginal. Third, increasing the number of associated APs leads to a higher capacity gain for the C-MC strategy, but a lower or even negative capacity gain for the R-MC strategy. Thus, it may not be practical to use the ``lightweight'' R-MC strategy in THz communication systems, which is different from the conclusion drawn for mmWave communication systems. Fourth, the optimal density of APs that maximizes the capacity gain can be determined by using our analysis.

\bibliographystyle{IEEEtran}
\bibliography{ref1_2020ICC}

\end{document}